\documentclass[twocolumn,prl,showpacs]{revtex4}

\usepackage{graphicx}
\usepackage{amsmath}

\begin{document}

\title{Delay Induced Excitability}

\author{Tomasz Piwonski, John Houlihan, Thomas Busch and Guillaume Huyet}

\affiliation{ Physics Department, National University of Ireland,
                  University College Cork, Ireland}

\date{\today}


\begin{abstract}
  We analyse the stochastic dynamics of a bistable system under the
  influence of time-delayed feedback. Assuming an asymmetric
  potential, we show the existence of a regime in which the systems
  dynamic displays excitability by calculating the relevant residence
  time distributions and correlation times. Experimentally we then
  observe this behaviour in the polarization dynamics of a vertical
  cavity surface emitting laser with opto-electronic feedback.
  Extending these observations to two-dimensional systems with
  dispersive coupling we finally show numerically that delay induced
  excitability can lead to the appearance of propagating wave-fronts
  and spirals.
\end{abstract}

\maketitle 

The influence of time delayed feedback on the dynamics of otherwise
Markovian processes with noise is currently a very active topic of
research, as memory effects have been found to be important in many
systems in areas ranging from biology to medicine and electronics
\cite{Applications}.  Even though many of the general properties of
systems with memory are still unexplored, various methods for
engineering of the delay to induce instabilities or stabilize complex
dynamics have already been suggested \cite{Janson:04}.

As a generic model to study the effects of delay on a bi-stable
stochastic system one can consider a particle trapped in a double well
potential. In the absence of feedback the particle makes random
transitions between the two minima of the potential and two
consecutive jumps can be considered as being independed of each other.
When feedback is present, however, correlations are introduced into
the system and the dynamics becomes inherently non-Markovian. Let us
consider the following standard stochastic differential equation
\begin{equation}
  \dot x = -\frac{dU}{dx} + \epsilon x(t-\tau) + \eta (t),
  \label{eq:Langevin}
\end{equation}
which describes an over-damped motion in a double well potential
$U(x)$ under the influence of memory. The length of the delay
interval is $\tau$ and $\epsilon$ is the feedback strength.  Here
$\eta (t)$ describes Gaussian white noise with $\langle \eta (t)
\rangle =0$ and $\langle \eta (t) \eta (t')\rangle = 2 D \delta
(t-t')$. In the following we will denote the delayed state as
$x_\tau=x(t-\tau)$.

Without delay ($\epsilon=0$), the random switching between the two
minima of the potential, $x_{l,r}$, can be very well described by
Kramers' theory \cite{Kramers:40}. The residence time distribution
(RTD) $P_{l,r}(T)$ in each well is then given by
\begin{equation}
 P_{l,r}(T) =\gamma_{l,r}\; e^{-\gamma_{l,r}T},
\end{equation}
where the inverse of the switching rate is the so-called Kramers'
time, $T_{l,r}=1/\gamma_{l,r}$. It describes the average time between
two transitions
\begin{equation}
   T_{l,r}=\frac{2\pi}{\sqrt{U''(x_{l,r})U''(x_0)}}
           \exp\left[ \Delta U_{l,r} / D\right].
 \label{eq:KramersTime} 
\end{equation}
Here $x_0$ is the position of the local maximum between the two minima
$x_{l,r}$ and $\Delta U_{l,r} = U(x_0)-U(x_{l,r})$ is the potential
barrier for each well.

The statistical properties of eq.~\eqref{eq:Langevin} for
$\epsilon\neq 0$ have recently been analysed in the case of the
generic symmetric potential $U(x)=-x^2/2 +x^4/4$.  Noting that the
potential can be rewritten to include the delay term, $V(x)= U(x) +
\epsilon x\,x_\tau$, it can immediately be seen that the potential
barrier, and therefore the escape rates, become dependent on the state
at the earlier time $t-\tau$ \cite{Houlihan:04}. Hence, the continuous
system of eq.~\eqref{eq:Langevin} can be approximated by a discrete
two-state model, $r(t)=\pm 1$, with transition rates that depend on
the delayed state \cite{Tsimring:01}.  These rates depend on the
relative value of $r(t)$ and $r(t-\tau )$ and are defined as $p_1$ if
$r(t)=r(t-\tau )$ and $p_2$ if $r(t)\neq r(t-\tau)$ for a symmetric
potential.  Tsimring and Pikovsky have recently derived the power
spectrum for this model and shown the existence of coherence resonance
\cite{Tsimring:01}.  Other works have calculated and measured the
RTD \cite{Masoller:03,Houlihan:04,Curtin:04}.

\begin{figure}[b]
  \includegraphics[width=\linewidth]{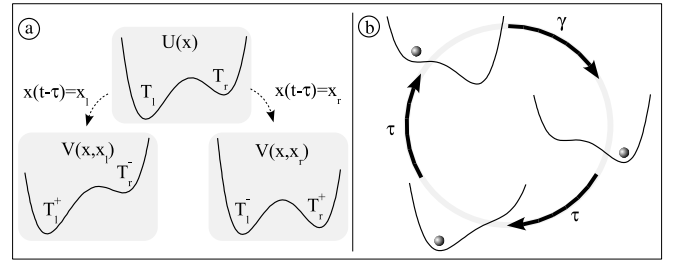}
  \caption{(a) Schematic definition of Kramers' residence times for an
    asymmetric potential without (upper graph) and with (lower graphs)
    negative feedback. (b) Schematic of the potentials during one
    excitable cycle.}
  \label{fig:Schematic}
\end{figure}

Here we show that the presence of an asymmetry in the potential can
lead to the appearance of excitability \cite{Lindner:04}.  In an
excitable system a perturbation above a certain threshold transfers an
otherwise stable steady-state (the resting state) into an excited
state (the firing state). The system then goes through a well defined
refractory cycle, before returning to its initial state.  During the
refractory cycle it is impervious to any external perturbation and
cannot 'fire' a second time. Systems following this kind of dynamics
are widespread in biology, chemistry and many other areas and there
are several mathematical models existing to describe such dynamics
\cite{Lindner:04}. It is worth pointing out, that for noise driven
excitable systems different manifestations of a higher degree of
order at a finite noise level have been discovered, for example
coherence resonance \cite{Pikovski:97} and wavefronts and even
self-ordered patterns like spirals in spatially extended systems
\cite{Ojalvo:99} .

For the asymmetric potential the presence of feedback leads to a
dependence of the amplitude of each potential barrier on the position
of the particle at the time $t-\tau$ and the dynamics has to be
described by four characteristic times $T_{l,r}^\pm$.  The times
$T_l^+$ and $T_l^-$ are the mean escape times from the left well when
the particle was in the left or right well at $t-\tau$, respectively,
and $T_r^-$ and $T_r^+$ are the mean escape times from the right well
if the particle was in the left or right well at $t-\tau$,
respectively (see Fig.~\ref{fig:Schematic}a).

Even though all transitions are inherently stochastic, we show below
that for negative feedback ($\epsilon < 0$) a parameter region exists,
in which the systems displays excitable behaviour. To illustrate this
let us consider the case of an asymmetric potential with $T_r\ll T_l$
(for $\epsilon=0$) such that the minimum corresponding to $x<0$ exists
for any value of the feedback, $x_\tau$, but the minimum for $x>0$
only exists for $x_\tau<0$ (see Fig.~\ref{fig:Schematic}b).  Let us
also assume that the particle has been in the left well for $t<0$ and
it is driven by noise into the right well at $t=0$. For small noise
amplitudes ($T_r^-\gg \tau$) the particle will almost certainly remain
in the right well until the value of the feedback term changes at
$t=\tau$.  The potential minimum on the right hand side then
disappears and the particle will move toward the remaining minimum at
$x<0$.  After a further time interval $\tau$, the potential will
switch to its initial bi-stable shape and the above process can start
over again.  Such a cycle is characteristic of an excitable system.
The initial state is stable under small perturbations but large
perturbations can induce a jump.  The firing corresponds to the
transition of the particle into the right well and the refractory time
consists of the two subsequent periods of duration $\tau$ before the
initial state is established again. During the first one the
refractory time the particle resides in the well $x>0$, from which
transitions are unlikely and during the second one only one minimum is
present and transitions cannot occur.

\begin{figure}
  \includegraphics[width=\linewidth]{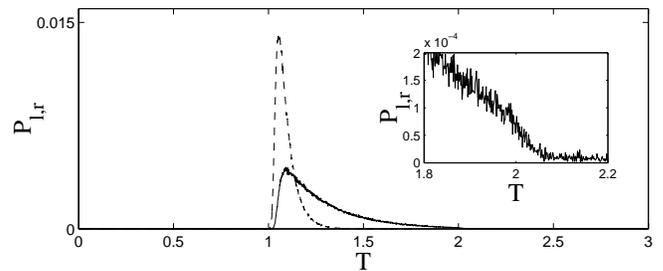}
  \caption{Numerical RTDs for the left (full line) and right well
    (dashed line). The delay interval is $\tau=1$. The inset shows the
    change in slope at $T=2$ for particles in the left well, which is
    due to a higher order effect in the feedback strength.}
\label{fig:RTD}
\end{figure}

To show that the dynamics described above can indeed be observed from
eq.~\eqref{eq:Langevin} we have numerically calculated the RTDs in the
excitable regime (see Fig.~\ref{fig:RTD}).  The potential was chosen
to be
\begin{equation}
  U(x)=-\frac{1}{2}x^2+\frac{1}{4}x^4+\frac{1}{3}\kappa x^3,
  \label{eq:AsymPot}  
\end{equation}
with $\kappa=0.3$ and $\epsilon=1/3$.  Assuming that the particle has
made a transition into the right well (the system has fired) at the
time $t=0$, one can clearly see from Fig.~\ref{fig:RTD} that the
transition probability back into the left well (dashed line) is
effectively zero in the interval $[0,\tau]$ The spike appearing for
$T>\tau$ then indicates the non-stochastic roll-over to the left well
once the feedback has changed and the potential possesses only the
single minimum on the left hand side.  From there no further
transition is possible until at a time $\tau$ later the potential
changes back to its bistable shape (full line in Fig.~\ref{fig:RTD}).
After that the distribution for $t>\tau$ follows Kramers' law.
Fig.~\ref{fig:RTD} clearly shows the existence of a refractory time
that forbids firing twice in an interval of 2$\tau$.

Let us compare our continuous system to a discrete model. Although the
simplest discrete model to describe excitability can be constructed
using a two-state system, it was recently shown, that networks of
excitable units display a wider spectrum of dynamical effects when
they are approximated by three state models
\cite{Nikitin:01,Prager:03}. In these models only the transition from
the resting into the firing state is stochastic, while the other two
transitions are deterministic with fixed waiting times. It
is clear from the discussion above that such a model is very close to
our bistable system with feedback.  Even though all transitions in the
potential are at all times inherently stochastic, the two intervals
during the cycle in which the transition probability is very low can
be well approximated by a waiting time of length $T_w=\tau$ (see
Fig.~\ref{fig:RTD}).

\begin{figure}
  \includegraphics[width=\linewidth]{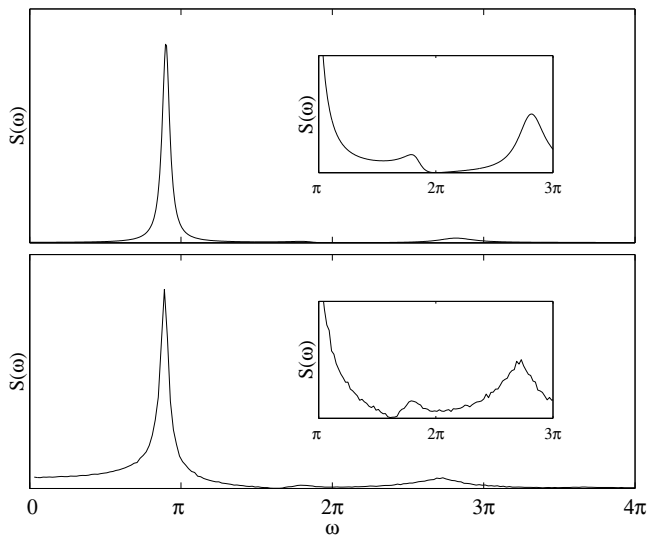}
  \caption{Power spectrum of the discrete (upper) and the
    continuous model (lower). Here $\tau=1$, $\epsilon=0.33$,
    $\kappa=0.3$.}
  \label{fig:PowerSpectrum}
\end{figure}

\begin{figure}[tb]
  \includegraphics[width=\linewidth]{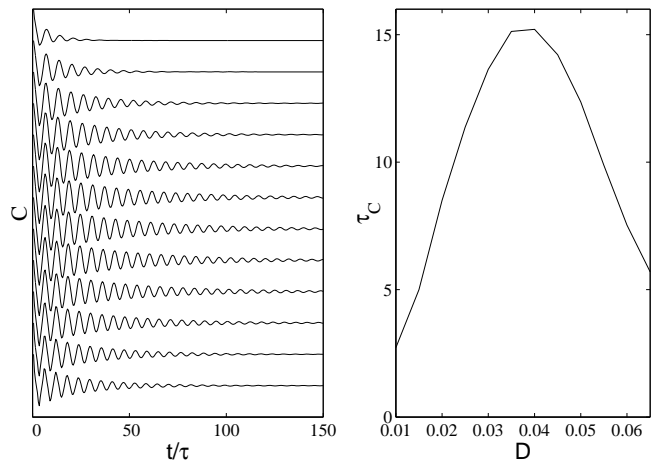}
  \caption{Left: autocorrelation function of the system with
    increasing noise. For the uppermost curve $D=0.01$ which increases
    downwards in steps of $\Delta D=0.005$. Right: correlation times.
    The maximum is a clear sign of ordering.}
\label{fig:CorrelationTime}
\end{figure}

To compare the power spectra we will assume that the stochastic output
of the discrete model is described by a function that has a value 1 if
the system is in the firing state and is zero otherwise. The spectrum
can then be calculated analytically \cite{Stratonovich:63}
\begin{equation}
  S(\omega)=\frac{2-2\cos(\omega\tau)}
                 {(\gamma+2\gamma^2\tau)(1-\cos(2\omega\tau)
                   +\frac{\omega}{\gamma}\sin(2\omega\tau)
                   +\frac{\omega^2}{2\gamma^2})}
  \label{eq:PowerSpectrum}  
\end{equation}
Fig.~\ref{fig:PowerSpectrum} shows the power spectrum given by
eq.~\eqref{eq:PowerSpectrum} (upper graph) and calculated from
numerical simulations of eq.~\eqref{eq:Langevin} with the potential of
eq.~\eqref{eq:AsymPot} (lower graph). Both curves are similar in shape
and show their maxima close to odd multiples of $\pi$ and minima at
even multiples of $\pi$.  However, it must be noted that for the
discrete model the coherence is an increasing function with noise
\cite{Nikitin:01,Prager:03}, while for the continuous model it is
maximized for a finite strength of the noise.  To demonstrate this
resonance we have calculated the autocorrelation function and the
associated correlation times, $\tau_C$, as a function of noise
strength. The correlation time is defined as the integral over the
square of the autocorrelation function \cite{Pikovski:97}
\begin{equation}
  T_c=\int_0^\infty \left[\frac{\langle\tilde x(t)\tilde x(t+t^\prime)\rangle}
                {\langle\tilde x(t)^2\rangle}\right]^2dt^\prime,\qquad
   \tilde x=x-\langle x\rangle\;.
\end{equation}
The left hand side of Fig.~\ref{fig:CorrelationTime} shows the
autocorrelation functions with noise increasing from the top to the
bottom graph. It is clearly visible that the correlations are
strongest for a finite noise level, which is confirmed in the plot on
the right hand side, where the $\tau_C$ show a clear maximum for a
noise amplitude of $D\sim 0.04$. This behaviour is closely related to
the appearance of a phase transition from excitable to oscillatory
media  \cite{Alonso:01,Ullner:03}.

\section{Experimental Results}
\label{sec:VCSEL}

Excitability has been observed in the intensity of semiconductor
lasers with optical feedback \cite{Giudici:97}. Here we experimentally
analyse the behaviour of the polarisation dynamics of a Vertical
Cavity Surface Emitting Laser (VCSEL) with opto-electronic feedback as
described in \cite{Houlihan:04,Curtin:04}.  The light emitted by a
VCSEL is usually linearly polarised but, as the injection current
increases, the emission axis will switch between the two orthogonal
states. Around this switching point one can find a current range in
which the light polarisation randomly switches between the two axes.
Previous experiments have been designed using these lasers to study
Kramers' law \cite{Willemsen:98}, stochastic resonance
\cite{Giacomelli:98} and the properties of noisy time delay dynamical
systems \cite{Houlihan:04,Curtin:04}. It is worthwhile noticing that
the shape of the potential depends on the bias current applied to the
laser. At the center of the bistable region both polarisations occur
with the same probability and the associated potential is symmetric.
The potential becomes asymmetric as the injection current is varied
toward the boundary of the the bistable region and our laser was
operated in this region in order to observe delay-induced
excitability.

\begin{figure}[htb]
\includegraphics[width=\linewidth]{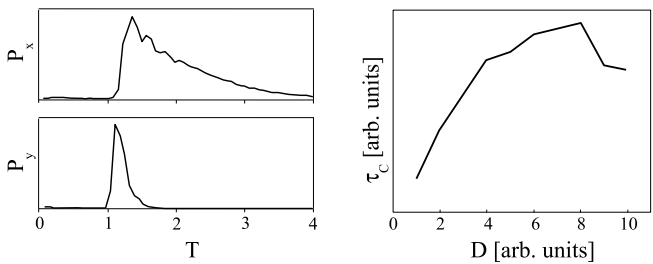}
\caption{Left: Experimental RTDs for the $x$ and $y$ polarization of
  the VCSEL. Excitability is clearly visible from absence of residence
  times shorter than $\tau$. Right: Correlation times of the VCSEL
  output signal for increasing noise levels.}
\label{fig:RTDexp}
\end{figure}

Although spontaneous polarisation switching was observed without the
addition of external noise, a noise generator was added in order to
observe coherence resonance.  Fig.~\ref{fig:RTDexp} (left) shows the
measured RTDs for the lowest noise in the excitable regime where
$P_{x},P_{y}$ refer to the orthogonal linear polarisation states. The
important feature is the absence of switching events for times less
than $\tau$.  Additionally, both the stochastic nature of the firing
transition and deterministic nature of the refractory cycle can be
seen in the spread of switching times being much less for $P_{y}$ than
$P_{x}$.  Coherence resonance for this operating point is shown in
Fig.~\ref{fig:RTDexp} (right) and qualitatively reproduces the
predicted numerical behaviour.

\section{Two-Dimensional Systems}
\label{sec:2D}

\begin{figure}[htb]
  \includegraphics[width=\linewidth]{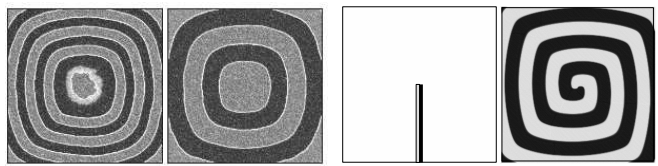}
  \caption{Left: Two dimensional wave-surfaces. The delay time in the
    left panel is twice the value of the delay time in the right
    panel. Right: Spiral pattern emerging from the initial condition
    (left panel) as described in the text.  Black corresponds to the
    system in the firing state, grey to the refractory state and white
    to the normal state. The parameters here are $\tau=80$,
    $\epsilon=0.343$, $\kappa=0.07$ and $D=0$.}
\label{fig:WaveFronts}
\end{figure}

The ability to sustain spatio-temporal patterns in systems of coupled
stochastic units is an important property of excitable systems
\cite{Ojalvo:99}. It was shown that wave-propagation and spiral
patterns \cite{Winfree:72} are common in excitable systems and here we
confirm their existence for spatially coupled elements as described by
eq.~\eqref{eq:Langevin}.  We consider a situations where the elements
are coupled via a diffusion term
\begin{equation}
    \frac{\partial s}{\partial t}=s-\kappa s^2-s^3-\epsilon s_\tau 
            +\sqrt{2D}\xi +\nabla^2 s\;.
\end{equation}
Let us choose each element of the system to be initially in the
resting state. At some point the noise will trigger one of the
elements to fire and the coupling to neighboring elements will lead to
short time correlations: in order to reduce the 'kinetic energy'
between two neighbouring elements they will also undergo a transition
into the firing state. After the refractory time the elements will
transfer back in the initial state, marking the end of the wave-front.
However, due to higher order effects in the feedback strength
\cite{Curtin:04}, the states at the end of the refractory time do not
have the same Kramers' time as the initial ones. In fact, for negative
feedback the new Kramers' time is much smaller and these elements are
more prone to being excited again. This leads to the continuous
creation of wave surfaces with wavelength $\tau$.  In
Fig.~\ref{fig:WaveFronts} (left) we show two examples of such wave
surfaces, where the delay time of the system on the right is twice as
long as in the system shown on the left hand side. The two dominating
colors represent the two refractory states for $s>0$ and $s<0$.  It is
clear that these ordered waves can only be sustained at an optimal
noise level.  Too little noise will not lead to any excitations
whereas too much noise will lead to fragmentation of the wavefronts.

Instead of using noise to trigger the spatial ordering, one can also
choose appropriate initial conditions. Here we demonstrate the
appearance of spiral patterns by preparing a system with one area
where the elements have just fired after having been in the resting
state for $[-\tau,0[$~. A neighboring area has just made the
transition back into the resting state, while having been in the
refractory state for $[-\tau,0[$. The rest of the system is in the
resting state and has been there for $[-\tau,0]$ (see
Fig.~\ref{fig:WaveFronts} (right)).  Integrating from these intial
conditions leads to the evolution of a spiral as shown on the right
hand side in Fig.~\ref{fig:WaveFronts}.

In summary we have shown that including delay into bistable systems
can lead to excitable behaviour. We have theoretically investigated
the model of an asymmetric double well potentials and found clear
signatures of excitability in the RTD. We have also compared the
continuous to a discrete model for excitability and found good
agreement of the power spectra.  We have calculated the coherence
times and shown the existence of coherence resonance.  Experimentally
we have demonstrated the above behaviour using a VCSEL in the
bi-stable regime.For spatially extended coupled systems we have
demonstrated the appearance of spatio-temporal order in the form
of waves and spirals.

\acknowledgments

This work was supported by Science Foundation Ireland under contract
number sfi/01/fi/co.


\begin{thebibliography}{}

\bibitem{Applications}
% Noise and critical behavior of the pupil light reflex at oscillation
% onset
A.~Longtin, J.G.~Milton, J.E.~Bos, and M.C.~Mackey, Phys.~Rev.~A {\bf
  41}, 6992 (1990); 
% Long Memory Processes ( 1/f alpha  Type) in Human Coordination
Y.~Chen, M.~Ding, and J.A..S.~Kelso, Phys.~Rev.~Lett.~{\bf 79}, 4501
(1997);
% Effect of external noise correlation in optical coherence resonance
J.M.~Buld\'u, J.~García-Ojalvo, C.R.~Mirasso, M.C.~Torrent, and
J.M.~Sancho, Phys.~Rev.~E {\bf 64}, 051109 (2001);
% Stochasticity in Transcriptional Regulation: Origins, Consequences,
% and Mathematical Representations
T.B.~Kepler and T.C.~Elston, Biophys.~J.~{\bf 81}, 3116 (2001).

% Delayed Feedback as a Means of Control of Noise-Induced Motion
\bibitem{Janson:04} N.B.~Janson, A.G.~Balanov, and E.~Sch\"oll,
  Phys.~Rev.~Lett.~{\bf 93}, 010601 (2004).

%
\bibitem{Kramers:40} H. Kramers, Physica (Utrecht) \textbf{7}, 284 (1940).

% Experimental investigation of a bistable system in the presence of noise 
% and delay  
\bibitem{Houlihan:04} J.~Houlihan, D.~Goulding, Th.~Busch, C.~Masoller
  and G.~Huyet, Phys.~Rev.~Lett.~{\bf 92}, 050601 (2004).

% Noise-Induced Dynamics in Bistable Systems with Delay
\bibitem{Tsimring:01} L.S.~Tsimring and A.~Pikovsky,
  Phys.~Rev.~Lett.~{\bf 87}, 250602 (2001).

% Distribution of Residence Times of Time-Delayed Bistable Systems 
% Driven by Noise
\bibitem{Masoller:03} C.~Masoller, Phys.~Rev.~Lett.~\textbf{90},
  020601 (2003).

% Distribution of Residence Times in Bistable Noisy Systems with 
% Time-Delayed Feedback 
\bibitem{Curtin:04} D.~Curtin, S.P.~Hegarty, D.~Goulding, J.~Houlihan,
  Th.~Busch, C.~Masoller and G.~Huyet, Phys.~Rev.~E \textbf{70},
  031103 (2004).

% Effects of noise in excitable systems
\bibitem{Lindner:04} B.~Lindner, J.~Garc{\'i}a-Ojalvo, A.~Neiman, and
  L.~Schimansky-Geier, Phys.~Rep~{\bf 392}, 321 (2004).

% Coherence resonance in a noise-driven excitable system
 \bibitem{Pikovski:97} A.S.~Pikovsky and J.~Kurths,
   Phys.~Rev.~Lett.~\textbf{78}, 775 (1997).

% Noise in Spatially Extended Systems
\bibitem{Ojalvo:99} J.~Garc{\'i}a-Ojalvo, J.M.~Sancho, Noise in
  Spatially Extended Systems, Springer, New York, 1999. 

% Collective Dynamics of Two-Mode Stochastic Oscillators
\bibitem{Nikitin:01} A.~\~Nikitin, Z.~N\' eda, and T.~Vicsek,
  Phys.~Rev.~Lett.~{\bf 87}, 024101 (2001).

% Coupled three-state oscillators
\bibitem{Prager:03} T.~Prager, B.~Naundorf, and L.~Schimansky-Geier,
  Physica A {\bf 325}, 176 (2003).

% Regular Wave Propagation Out of Noise in Chemical Active Media
\bibitem{Alonso:01} S.~Alonso, I. Sendi\~na-Nadal,
  V.~P\'erez-Mu\~nuzuri, J.M.~Sancho, and F.~Sagu\'es,
  Phys.~Rev.~Lett.~{\bf 87}, 078302 (2001).

% Noise-Induced Excitability in Oscillatory Media
\bibitem{Ullner:03} E.~Ullner, A.~Zaikin, J.~Garc\'ia-Ojalvo, and
  J.~Kurths, Phys.~Rev.~Lett.~{\bf 91}, 180601 (2003).

% Polarization Switching of a Vertical-Cavity Semiconductor Laser as a 
% Kramers Hopping Problem
\bibitem{Willemsen:98} M.B.~Willemsen, M.U.F.~Khalid, M.P.~van Exter,
  and J.P.~Woerdman , Phys.~Rev.~Lett. {\bf 82}, 4815 (1999).

% Andronov bifurcation and excitability in semiconductor lasers with 
% optical feedback
\bibitem{Giudici:97} M.~Giudici, C.~Green, G.~Giacomelli, U.~Nespolo,
  and J.R.~Tredicce, Phys.~Rev.~E {\bf 55}, 6414 (1997).

% Statistics of Polarization Competition in VCSELs
\bibitem{Giacomelli:98} G.~Giacomelli and F.~Marin, Quantum
  Semiclassic.~Opt. {\bf 10}, 469 (1998).
% Stochastic resonance in vertical cavity surface emitting lasers
  S.~Barbay, G.~Giacomelli and F.~Marin, Phys.~Rev.~E {\bf 61}, 157
   (2000).


% Topics in the Theory of Random Noise
\bibitem{Stratonovich:63} R.L.~Stratonovich, {\sl Topics in the Theory of
  Random Noise} (Gordon and Breach, New York, 1963).

%
\bibitem{Winfree:72} A.T.~Winfree, Science {\bf 240}, 460 (1972).


\end{thebibliography}
\end{document}